\documentclass[preprint, journal]{vgtc}             

\usepackage{enumitem}
\usepackage{amsmath}
\usepackage[pagebackref,bookmarks]{hyperref}  
\usepackage{multirow}
\usepackage{graphicx}
\usepackage{tabu}                      
\usepackage{booktabs}                  
\usepackage{lipsum}                    
\usepackage{mwe}                       
\usepackage{flushend}
\usepackage{mathptmx}                  

\newcommand{\SM}{\textbf{\texttt{SM}}\xspace}
\newcommand{\DM}{\textbf{\texttt{DM}}\xspace}
\newcommand{\SSelect}{\textbf{\texttt{SS}}\xspace}
\newcommand{\DS}{\textbf{\texttt{DS}}\xspace}

\newcommand{\Symbolic}{\textbf{\texttt{S}}\xspace}
\newcommand{\Direct}{\textbf{\texttt{D}}\xspace}
\newcommand{\Select}{\textbf{\texttt{S}}\xspace}
\newcommand{\Manipulate}{\textbf{\texttt{M}}\xspace}

\renewcommand{\figureautorefname}{Figure}

\preprinttext{To Appear in 2025 TVCG Special Issue on the 2025 IEEE Conference on Virtual Reality and 3D User Interfaces (VR)}

\vgtccategory{Research}

\vgtcpapertype{theory/model}

\title{Tap into Reality: Understanding the Impact of Interactions on Presence and Reaction Time in Mixed Reality}

\author{Yasra Chandio\thanks{e-mail: ychandio@umass.edu}\\ %
        \scriptsize University of Massachusetts Amherst\\
\and  Victoria Interrante\thanks{e-mail: interran@umn.edu}\\ %
     \scriptsize University of Minnesota Twin Cities %
\and Fatima M. Anwar \thanks{e-mail: fanwar@umass.edu}\\ %
     \scriptsize University of Massachusetts Amherst %
     }

\abstract{%
Enhancing presence in mixed reality (MR) relies on precise measurement and quantification. While presence has traditionally been measured through subjective questionnaires, recent research links presence with objective metrics like reaction time. Past studies examined this correlation with varying technical factors (object realism and behavior) and human conditioning, but the impact of interaction remains unclear.
To answer this question, we conducted a within-subjects study ($N=50$) to explore the correlation between presence and reaction time across two interaction scenarios (\emph{direct} and \emph{symbolic}) with two tasks (\emph{selection} and \emph{manipulation}). We found that presence scores and reaction times are correlated (correlation coefficient of $-0.54$), suggesting that the impact of interaction on reaction time correlates with its effect on presence. 
}

\keywords{Mixed Reality, Presence, Interaction}

\teaser{
    \centering
    \begin{tabular}{c}
    \includegraphics[width=\textwidth]{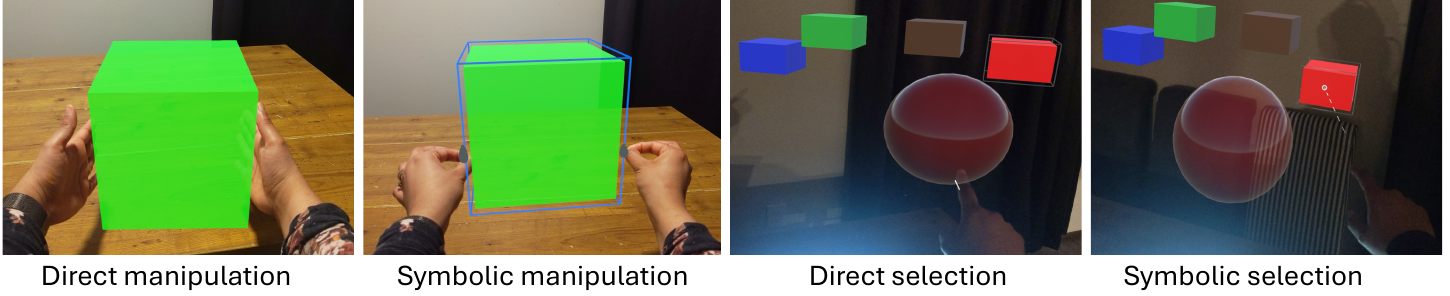} 
    \end{tabular}
    \vspace{-0.3cm}
    \caption{View of the virtual scene for four conditions in MR. In the direct manipulation (\DM) condition (left), a box is grabbed with both hands. An unnatural two-hand grab gesture with a hook is used in the symbolic manipulation (\SM) condition (center-left). For the direct selection (\DS) condition (center-right), a bubble is popped by touch, while in the symbolic selection (\SSelect) condition (right), an air-tap with the ray is used to interact with the box.
 }
  \label{fig:teaser}
}

\nocopyrightspace

\begin{document}

\firstsection{Introduction}
\label{sec:introduction}
\maketitle

In Mixed Reality (MR), creating a sense of presence is essential to providing an immersive and engaging experience. Presence, the psychological sensation of \emph{being there}, allows users to feel fully absorbed in a virtual environment (VE)~\cite{presence-1}. Presence can take various forms, including spatial presence (the feeling of physically existing in the VE)~\cite{presence-2}, social presence (the perception of interaction with others)~\cite{presence-social-presence}, and self-presence (the sense of embodiment and identity within the VE)~\cite{presence-measure-subjective-objective,ipq}. This study measures spatial presence, aligning with the tasks designed to assess user interactions in VE. Measuring presence is essential for improving it, and traditional approaches have primarily relied on subjective questionnaires. However, recent research has identified correlations between presence and objective measures, such as reaction time~\cite{chandio-tvcg-23,chandio-vr-24-human-factors}. These correlations have been mainly explored when presence is influenced by technical factors such as the appearance and behavior of virtual elements
~\cite{ws, Place-illusion-plausibility} or by conditioning-related human factors~\cite{cognition-perception-Latoschik_2022, plausible-tvcg-23}.

While these findings have provided valuable insights into the nature of presence, prior work often assumes that specific interaction mechanisms do not affect the relationship between presence and reaction time. However, literature~\cite{van2000interaction-presence-2000,muender2022haptic-fidelity,bimanual-interactionhough2015fidelity,ux-social-interactivity-vergari2021influence,lok2003effects-interaction, mcmahan2012evaluating}, has suggested that interaction design itself influences presence. 
In MR, \emph{interaction refers to the ways in which users engage and manipulate the VE}. It encompasses the mechanisms and interfaces that allow users to perceive, navigate, and affect the virtual world. Biocca's work~\cite{biocca-interaction} on interactions in VE suggests that users build a mental model of interaction patterns, and any deviation from expected behavior can affect presence. 
Similarly, expectations about the realism and consistency of interactions are shown to be essential in establishing a presence in MR environments~\cite{perception-emotion-expectation-paes2021evidence, expectation-presnce-ling2013relationship,cognition-perception-Latoschik_2022}. For example, natural interactions may lead to greater presence, while unfamiliar or inconsistent interactions can disrupt presence~\cite{FIFA-mcmahan2016interaction, table-FIFA-shafer2021effects}.
On the other hand, \emph{reaction time is typically measured as a response to a cue}, often occurring as an interaction, and has been shown to vary based on scene fidelity~\cite{chandio-tvcg-23} and user conditioning~\cite{chandio-vr-24-human-factors}. Therefore, any approach to measuring presence should be sensitive to interactions to provide accurate experience assessments, as it directly influences reaction time and impacts presence~\cite{ipq}. 

To address this gap in the literature, we conducted an exploratory within-subjects study ($N=50$). We evaluated the relationship between presence and reaction time across two interaction mechanisms: \emph{direct (\Direct)} and \emph{symbolic (\Symbolic)}. 
Direct interaction closely mimics real-world actions, such as reaching out to grab or touch a virtual object. Symbolic interaction, on the other hand, often employs abstract mechanisms such as pressing a button or performing a gesture that does not replicate real-world actions but achieves the intended effect. Still, symbolic gestures may also involve degrees of familiarity, depending on user conditioning and task goals~\cite{interaction-mental-model}.
We examined two tasks, \emph{manipulation} and \emph{selection}, equally amenable to \Direct and \Symbolic, to capture a broader range of interactions and to assess the joint effects of interaction type and task type. The manipulation task involves manipulating virtual objects directly, whereas the selection task demands precise coordination between user intentions and the virtual object.  
We followed a 2 (interaction: direct vs. symbolic) $\times$ 2 ( tasks: manipulation vs. selection) study design to create four conditions: \emph{direct manipulate} (\DM),  \emph{symbolic manipulate} (\SM), \emph{direct select} (\DS), \emph{symbolic select} (\SSelect). We used post-experience questionnaires to measure the change in presence between conditions and systematically measured the reaction time of users in response to a visual stimulus. The goal was to understand how these choices influenced presence and reaction time. Based on the collected data, we aimed to answer the following research questions (\textbf{RQ}):

\begin{enumerate}[leftmargin=0.83cm, itemsep=-0.1cm, topsep=-0.1cm]
    \item[\textbf{RQ1:}] To what extent does interaction type influence presence in MR?
    \item[\textbf{RQ2:}]How does the task type modulate the impact of interaction type on presence?
    \item[\textbf{RQ3:}] Does the correlation between presence and reaction time hold when the interaction is manipulated to alter presence? 
\end{enumerate}

\section{Background and Related Work}
\label{sec:background}
\subsection{Measuring Presence in MR}
Our definition of MR aligns with Augmented Reality (AR) on Milgram's reality-virtuality spectrum~\cite{milgram1995augmented}, where virtual elements merge seamlessly with the physical world. The core of this experience is ``presence," the feeling of ``being there," which relies on interactions that reflect real-world behavior~\cite{presence-1}. To refine the analysis, this work focuses on spatial presence, which is primarily influenced by the user’s perception of their location and interaction within the VE. Self and social presence are outside the scope of this study but remain relevant for multi-user and embodiment-focused MR experiences~\cite{ar-presence-gandy2010experiences}. Accurately measuring \emph{presence} in any VE (AR/VR/MR) presents unique challenges. Traditional post-experience questionnaires~\cite{ws, ipq, ipq-2, grassini2020questionnaire-review, sus, questionnaire-chi-20, in-VR-questionarrie, questionarie-prior-experience}, often used to measure presence, capture retrospective impressions of sensory fidelity and engagement but can lack real-time insights and may suffer from memory biases~\cite{schwind-pinky}. Originally designed for fully immersive VR~\cite{presence-object-stevens2002putting}, questionnaires have been adapted for AR/MR~\cite{questionarrie-in-realities-usoh2000using, context} but still primarily assess peripheral factors rather than intrinsic presence~\cite{ar-presence-gandy2010experiences, presence-non-immersive,presence-object-stevens2002putting, AR-presence-Regenbrecht2021MeasuringPI,arobject-questionaire-stevens2000sense}. Despite limitations, they are useful across the reality-virtuality spectrum~\cite{presence-mr} if all users experience the same VE.

Given the limitations of subjective measures, researchers have studied behavioral and physiological metrics like facial expressions~\cite{behivor-measure}, posture\cite{postural-response-freeman2000using}, heart rate~\cite{change-in-heart-rate}, and skin conductance~\cite{physiological-measure} as objective measures for presence. Still, these often yield inconsistent results~\cite{not-my-hands, schwind-pinky, in-VR-questionarrie}. Objective metrics, such as reaction time have shown potential in measuring presence~\cite{lombard1997heart, sus, sense-of-presence-barfield1993sense, basdogan2000experimental, perfromance-presence, ijsselsteijn2004presence, human-performance-1}. Recent research by Chandio et al.~\cite{chandio-tvcg-23, chandio-vr-24-human-factors} has demonstrated a correlation between presence and reaction time in MR environments: as presence decreases, reaction time increases. Particularly when influenced by technical elements such as place and plausibility illusions~\cite{ws, Place-illusion-plausibility} or by conditioning-related human factors, including cognitive perception and prior experiences~\cite{plausible-tvcg-23, cognition-perception-Latoschik_2022}.

\subsection{Presence and Interaction}
Despite evidence suggesting that interaction design influences presence, research on the impact of interaction has yielded mixed results. While some studies found that high-fidelity interactions enhance immersion and enjoyment~\cite{van2000interaction-presence-2000,muender2022haptic-fidelity,bimanual-interactionhough2015fidelity,lok2003effects-interaction, mcmahan2012evaluating,interaction-presence-shafer2024effect}, others reported similar outcomes between high- and low-fidelity interactions~\cite{questionnaire-chi-20,ux-social-interactivity-vergari2021influence}. 
Others have either focused on the sensor side of interaction~\cite{muender2022haptic-fidelity} or explored interaction modalities~\cite{luong2023controllersvshand}, but none have directly addressed the impact of presence. Many of these studies have only examined the interaction effect incidentally, treating it as a secondary research question or control measure rather than a central focus.
These inconsistencies underscore the need for dedicated research to comprehensively understand the relationship between interaction and presence.

\subsection{Interaction, Presence and Reaction time}
While research has explored technical and human factors influencing presence and its relationship with reaction time, the direct link between presence, interaction, and reaction time requires more attention. Interaction plays a vital role in shaping how presence and reaction time relate. Interaction is crucial because it engages the user and directly triggers their reaction to cues. Users engage with VE through different interaction mechanisms like controllers, hand gestures, or voice commands, and these interactions significantly affect their sense of presence and responsiveness~\cite{highlevel-coginition-reaction-time}.
For instance, familiar tasks like manipulating and selecting objects foster a sense of control and agency, helping users engage with the virtual world naturally. Immediate and consistent feedback through visual, auditory, or haptic cues reinforces user actions and strengthens the presence~\cite{muender2022haptic-fidelity}. However, the relationship between interaction, presence, and reaction time is challenging. Overly complex or unfamiliar controls can slow reaction time and reduce the sense of presence~\cite{expectation-presnce-ling2013relationship}. Furthermore, task familiarity is crucial while designing the interaction~\cite{perception-kemeny2003evaluating,improving-interaction-bottone2016improving}, as daily activities like moving and pointing need to be intuitive and consistent with users' mental models to enhance presence~\cite{murphy2020we-mental-model-expectation, interaction-mental-model} but may add latency to the reaction time.

\subsection{Contributions Beyond Related Work}
In this section, we highlight the key contributions of our work that extend beyond the existing literature, as follows:
\begin{enumerate} [leftmargin=0.5cm, itemsep=-0.1cm, topsep=-0.1cm]
\item \textbf{Interaction, presence, and reaction time model}:
We introduce a structured model that connects interaction types, tasks, presence, and reaction time. This model provides a framework for understanding how interaction fidelity influences presence and user reaction times.

\item \textbf{Generalizing the presence reaction time relationship} 
Existing research has suggested a moderate correlation between presence and reaction time, influenced by various technical and human factors. However, our study aims to extend this understanding in two crucial ways: (a) We test whether the established relationship between presence and reaction time applies to different scene elements. It is essential to assess the sensitivity of this relationship and its relevance across scene elements.
(b) Since reaction time is inherently captured through user interaction, examining this relationship in the context of specific interactions is critical. Our study focuses on how different interactions influence this relationship, providing insights into the interaction dynamics and their impact on the user. By addressing these aspects, our work contributes to a more comprehensive understanding of presence and reaction time in MR. 

\item \textbf{Task-Specific Analysis of Interaction Effects:}
Previous studies have not explored in detail how specific tasks interact with different interaction types to affect presence and reaction time. Our study complements existing findings by analyzing the impact of these particular tasks, task-specific interactions, and their joint effects on user experience (presence) and performance behavior (reaction time). This task-specific focus provides a better understanding of how different interactions affect users depending on the nature of the task.
\end{enumerate}

\section{Approach}
\label{sec:approach}

\begin{figure}[t]
    \centering
    \includegraphics[width=\columnwidth]{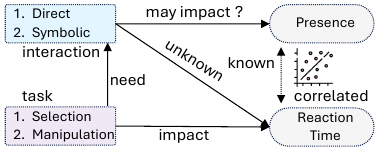}
    \vspace{-0.45cm}
    \caption{Model depicting the relationship between interaction types, tasks, presence, and reaction time.}
    \label{fig:rt_p_together}
    \vspace{-0.6cm}
\end{figure}

\subsection{Theoretical Model}
The correlation between presence and reaction time in MR, mainly when influenced by different interactions, presents an exciting area of research. In~\autoref{fig:rt_p_together}, we present a structured model that depicts how interaction types and tasks in MR affect presence and reaction time. Both of these are measurable quantities that may vary with different interaction-task combinations. The combination of these components forms the basis for the investigation of this paper. Now, we formally define the key elements of the model below:

\noindent \textbf{Presence.} is the psychological sensation of ``being there" within the VE~\cite{presence-mr}. 
Presence is about more than high-fidelity graphics; it requires a well-designed VE that replicates real-world experiences.

\noindent \textbf{Interactions.} are the methods and mechanisms through which users can manipulate virtual objects or engage with the VE. This includes direct physical hand gestures, voice commands, controller inputs, and eye tracking, which facilitate a communication loop between the user and the VE. 
A key factor in designing VE is determining how interaction fidelity\footnote{Interaction fidelity measures how accurately real-world interactions are replicated~\cite{mcmahan2012evaluating}.} supports high level goals of VE~\cite{interaction-performace-bhargava2018evaluating}. 
Interaction can be categorized in various ways~\cite{interaction-performace-bhargava2018evaluating, hand-tracking-jamalian2022effects, hand-manipulation-hololens2, interaction-presence-shafer2024effect}, such as single-handed versus two-handed or gesture-based versus controllers, near vs far interactions and so on. We focus on the high-level design of the interaction and classify that as follows:
\begin{itemize}[itemsep=-0.1cm, topsep=-0.1cm,leftmargin=0.35cm]
    \item \emph{Direct Interaction (\Direct)} is when the user engages with the VE in a manner that closely resembles real-world behavior. For instance, reaching out to grab a virtual object with a hand gesture, like grasping a physical object. 
    These interactions are considered more natural and may create a heightened sense of presence because they align with the user's natural, familiar movements and expectations.
    
   \item \emph{Symbolic Interaction (\Symbolic)} refers to actions that do not directly replicate real-world action but rely on symbolic or abstract representations to achieve the desired result. For instance, using a button to select a virtual object instead of touching it. Although less intuitive, these interactions are still effective in enabling users to interact with the VE. 
    Although \Direct interactions are preferred due to their realism~\cite{bergstrom2021evaluate-select-manipulate-interaction-survey}, they can be time-consuming and costly, requiring higher computational overhead. In contrast, \Symbolic interactions (raycasts or button presses) offer practical advantages in scenarios requiring fast, repetitive inputs, reduced physical effort, and interactions beyond arm's reach~\cite{interaction-performace-bhargava2018evaluating}.
\end{itemize}
\vspace{0.1cm}
\Direct is considered beneficial for precision-oriented tasks, such as assembly tasks that align with familiar motor behaviors~\cite{bowman2001introduction-interaction-survey}.
\Symbolic is practical for tasks needing less effort, and frequent actions, like sorting items via a button, are helpful when direct interaction is not feasible~\cite{bergstrom2021evaluate-select-manipulate-interaction-survey}. \Direct and \Symbolic are chosen for their foundational roles in MR, providing a solid base before extending research to other types. We limit our scope by avoiding broader interactions like voice, gaze, and haptics.

\noindent \textbf{Tasks.} represents a specific activity or goal users aim to accomplish in VE. It is an action-oriented step or process that the user undertakes while interacting with VE. 
Bowman et al.~\cite{bowman2001introduction-interaction-survey} categorized interaction techniques into selection, manipulation, travel, and system control tasks. LaViola et al.~\cite{laviola20173d-bowmen-survey-book} expanded on these by introducing canonical manipulation tasks: selection, positioning, rotation, and scaling. They highlighted the importance of using low-level tasks to evaluate interactions' impact. The following tasks are chosen based on their prevalence and relevance in everyday virtual interactions~\cite{bergstrom2021evaluate-select-manipulate-interaction-survey, koutsabasis2019empirical-mid-air-interaction}. 

\begin{itemize}[itemsep=-0.1cm, topsep=-0.1cm,leftmargin=0.35cm]
    \item \emph{Selection} tasks involve choosing or highlighting objects in the VE. For instance, clicking on or tapping an object to mark it as selected or initiating an actuation. The simplicity of the task usually means it requires less effort and physical demand.
    \item \emph{Manipulation} is a more intricate task that requires reaching out and manipulating virtual objects, such as grabbing, moving, or rotating items in VE. This type of task demands higher interaction fidelity than selection tasks due to the complexity of the movements involved, which makes it important for our evaluation.
\end{itemize}

These tasks are emphasized for their relevance in VE and their importance in assessing interaction impact. The aim was to choose tasks that were equally amenable to \Direct and \Symbolic without obvious prejudice towards either. Being equivalently easy to accomplish via either method allows us to explore the extent to which reaction time may be affected by differences in presence. 
The relationship between presence and reaction time in MR may be influenced by interactions and tasks in unexplored ways. What is known and unknown in this construct motivates further investigation.

\noindent \textbf{Known.} Some form of user interaction is required to complete any task in the VE. Each task has a measurable reaction time that provides insights into user performance.
Prior research has established a proven correlation between presence and reaction time~\cite{reaction-time-correaltion, interaction-fidelity-chi}.
Literature also supports the idea that interactions may affect presence~\cite{interaction-presence-shafer2024effect}. For instance, direct interactions that mimic real-world behavior can enhance presence, while symbolic or abstract interactions may be less effective because the user does not have an existing mental model to perform these symbolic interactions~\cite{interaction-mental-model}.

\noindent \textbf{Unknown.} It remains unclear precisely how different interaction types impact reaction times.
Further exploration is required to determine how much interaction influences cognitive and motor responses, resulting in faster or slower reaction times.
The connection between specific combinations of interaction types and task types and how they impact the correlation between presence and reaction time is yet to be fully understood.

\begin{figure}
    \centering
    \includegraphics[width=\columnwidth]{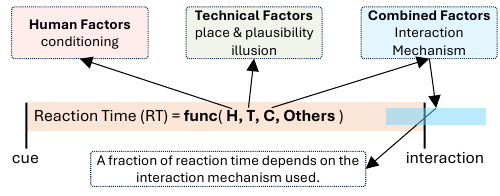}
    \vspace{-0.4cm}
    \caption{Reaction Time Decomposed.}
    \label{fig:rt_decomposed}
    \vspace{-0.3cm}
\end{figure}

\noindent \textbf{Factors affecting Reaction Time. }Figure\ref{fig:rt_decomposed} aims to break down the factors influencing reaction time (RT) in VE to design the study to learn the sensitivity of the relationship between presence and RT to interaction. The main focus of Figure\ref{fig:rt_decomposed} is explaining the different RT contributors. Technical factors represent the quality and realism of VE and impact how quickly users react, as more realistic environments can reduce the mental load on users~\cite{reaction-time-cognition}.
Human Factors are attributes that relate directly to the user and their physiological and psychological state. For example, the user's familiarity and prior experience with the VE affect their ability to respond quickly and effectively to stimuli. 
Combined factors consider the interactions between the human and technical aspects, such as interaction mechanisms. This affects how naturally a user can engage with VE; some mechanisms might require complex gestures or symbolic inputs that could slow down reaction times. The effectiveness of the interaction mechanism is crucial, as it directly contributes to how the VE appears and behaves and how the user will respond. Users build a mental model of interaction patterns based on expectations and past experiences~\cite{murphy2020we-mental-model-expectation}. Any deviation from expected behavior can affect presence; for instance, natural interactions, such as grabbing objects with realistic hand gestures, may lead to greater presence. In contrast, awkward or inconsistent interactions may disrupt presence, making users more aware of the virtual world's artificial nature. Expectations around the realism and consistency of interactions are crucial for user presence in VE. When interaction aligns closely with expectations (\Direct), it fosters a sense of flow that allows users to interact intuitively with the VE. Conversely, \Symbolic creates a mismatch that may reduce presence. Lastly, other factors may also influence RT, such as task complexity, cognitive load, and environmental distractions; exploration of these elements is out of the scope of this paper.
Due to interaction's potential influence, investigating reaction time for a given task and interaction pair and its variation is essential to understanding the correlation between presence and reaction time.

\subsection{Hypothesis}
\label{sec:hyp}
Given these observations, we hypothesize that the interaction mechanisms in MR can affect the presence and their reaction times to stimuli. 
Our hypotheses (\textbf{H}) concerning interactions are as follows: 
\begin{itemize}
[itemsep=-0.1cm, topsep=-0.1cm,leftmargin=0.35cm]
\item \textbf{H1:} Manipulating interactions (direct and symbolic) leads to a change in the presence.
\begin{itemize}[itemsep=-0.1cm, topsep=-0.1cm,leftmargin=0.35cm]
    \item \textbf{H1.1}: Participants experience a higher sense of presence during direct interactions than symbolic interactions.
\end{itemize}
\item \textbf{H2:} 
The effect of interaction type on presence is moderated by the type of task the user performs.  
\begin{itemize} [itemsep=-0.1cm, topsep=-0.1cm,leftmargin=0.35cm]
\item \textbf{H2.1}: The change in interaction type leads to a change in presence, irrespective of the type of task. 
\end{itemize}
\item \textbf{H3}: Presence and reaction time are correlated.
\begin{itemize} [itemsep=-0.1cm, topsep=-0.1cm,leftmargin=0.35cm]
\item \textbf{H3.1}: The change in interaction type leads to a change in reaction time, irrespective of the type of task. 
\item \textbf{H3.2}: The extent of the change in reaction time depends on the interaction type.
\end{itemize}
\end{itemize}
 
\subsection{Experimental Design}
\label{sec:exp-design}
\begin{figure}
    \centering
\includegraphics[width=\linewidth]{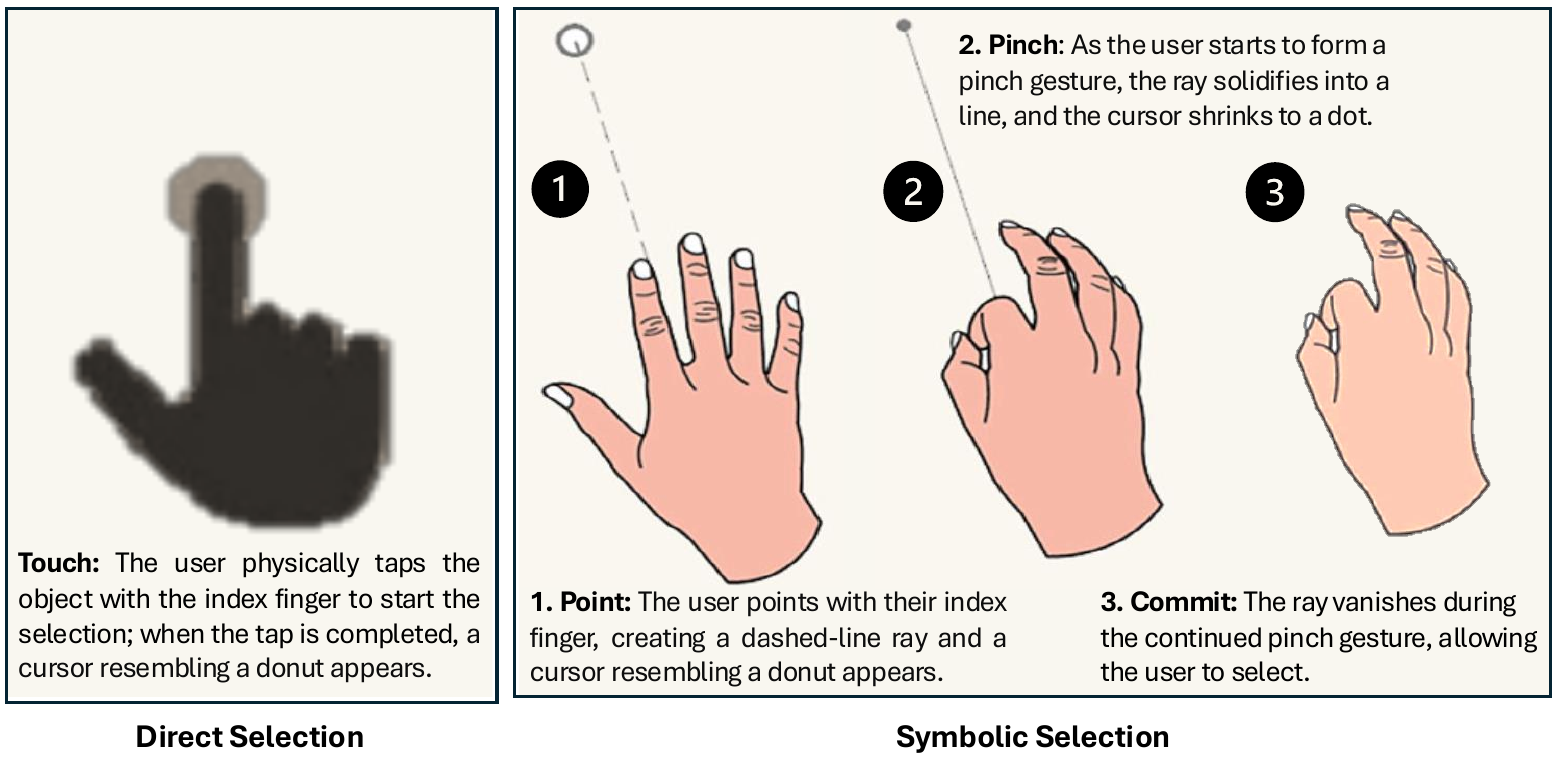}
    \caption{Hand gestures for selection task}
    \label{fig:select-task}
\end{figure}
\noindent \textbf{Selection Task} is designed as a color-matching bubble pop activity, where participants are required to match a bubble's color to a highlighted box among four options (red, green, blue, brown). We created a sequence of four color patterns, using a Latin square design to order the bubble colors repeatedly throughout the task. To determine which box to highlight, we employed the reverse diagonal values from our four color patterns matrix, ensuring that each color box was highlighted systematically and predictably across multiple trials.
In the direct selection (\DS) condition, participants are instructed to physically tap on the bubble directly with their index finger to pop it (make a selection as shown in \figureautorefname~\ref{fig:select-task}(left)) when its color-matched the highlighted box. Upon tapping, a donut-shaped cursor appears to confirm the selected object. This gesture replicates how users interact with bubbles in the real world.

In the symbolic selection (\SSelect) condition, when the color of the bubble matches the highlighted box, the participants are asked to perform the airtap interaction~\cite{air-tap} on the highlighted box to pop the bubble, as shown in \figureautorefname~\ref{fig:select-task}(right). 
The participants point toward the box using their index finger, which creates a ray depicted as a dashed line. A donut-shaped cursor appears on the box being pointed at, signifying where the participant is aiming. Then, the participants move their index finger toward their thumb, forming a pinch gesture.
As the participants transition to a pinch gesture, the dashed ray changes to a solid line, and the donut cursor shrinks to a dot, indicating that they are preparing to make a selection. Once the pinch gesture is sustained, the solid ray disappears entirely, and the participants can select the box. The selection is finalized when the pinch gesture is complete.
The placement of the bubble and boxes and interaction in \SSelect is carefully designed so that both the tasks in Select have the same physical workload and arm movement. Although the box is farther than the bubble, the intentional ray formation of the line of sight with the box compensates for the delta in the distance.
We use Fitts's law~\cite{fittslaw} to calculate the expected time of motor movement for both positions and corroborated that empirically during our pilot study (\S\ref{sec:pilot}).
In our design, the hypothesis assumption is that selection tasks benefit from direct touch interactions that may improve presence and are likely to reduce cognitive load. \Symbolic using an anchor/hook can be efficient for rapid selections despite some cognitive disconnect. Manipulation tasks, like grabbing and rotating, are more intuitive with \Direct and could increase the presence. \Symbolic interactions may require learning but can become efficient once mastered.
In the selection task, participants selected objects using either a button press (\Symbolic) or by reaching out to touch the target (\Direct). These differing interaction modes reflect the distinct mechanics of symbolic and direct interaction in MR. However, we recognize that the \Symbolic could have been implemented using a more consistent ray-cast targeting method rather than a button press. The choice of a button press was informed by its common usage in MR tasks involving symbolic actions but introduces a limitation due to the significant difference in selection targets between the two interaction types.

\begin{figure}[t]
    \centering
    \begin{tabular}{c}
    \includegraphics[width=\linewidth]{
    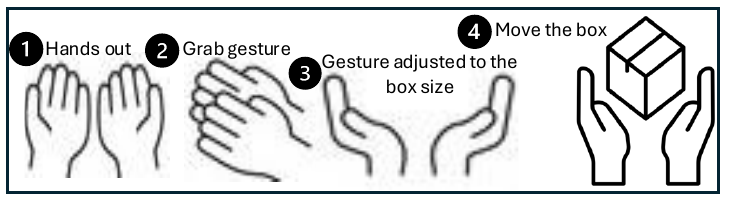} \vspace{-0.1cm}\\
    (a) Direct manipulation task\\
    \includegraphics[width=\linewidth]{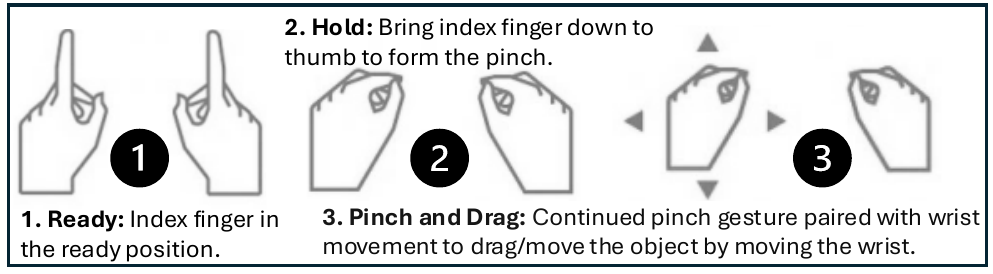} \vspace{-0.1cm} \\
    (b) Symbolic manipulation task \\
    \end{tabular}
    \vspace{-0.3cm}
    \caption{Hand gestures for the manipulation task.}
    \vspace{-0.4cm}
    \label{fig:manipulate-task}
\end{figure}

\noindent \textbf{Manipulation Task}
For the manipulation task, participants are presented with a virtual box positioned in front of them, which they must move back and forth each time the box turns green. In the direct manipulation (\DM) condition, participants performed a natural two-hand grab gesture to interact with the box, simulating real-life object movement, as shown in~\autoref{fig:manipulate-task}(a). Participants start by positioning their hands in front of them, palms facing up, and extending their fingers outward to prepare for the gesture.
Participants are then asked to flex their fingers inward, imitating a grab gesture. They adjust this gesture to match the box size, creating a realistic grip for the accurate interaction. Once a secure virtual grip was established, users could reposition the box by moving their wrists and arms. The box followed the hand movement, allowing users to relocate it as needed.

For the symbolic manipulation (\SM) condition, we implemented a bounding box around the virtual object. We highlighted a specific area (designed as a 'hook'), avoiding needing a full two-hand grab gesture. As illustrated in ~\autoref{fig:manipulate-task}(b), the participants hold their index finger in a pointing position to the hooks on each side of the box, slightly elevated and ready for action. This starting posture enables the system's accurate hand tracking. The participants move their index fingers toward their thumb, forming a pinch gesture. The thumb and index finger approach each other but do not necessarily make contact to indicate the intent to interact with a box—the hold phase transitions from mere pointing to active interaction. Once the pinch gesture is initiated, the participants maintain it while moving their wrists, effectively dragging/moving the box in the desired direction (back and forth).

\subsection{Interaction Implementation}
\vspace{-0.1cm}
Interactions in this study are managed through the HoloLens' built-in hand tracking (accuracy of Hololens2 hand tracking is empirically tested by~\cite{kern2021using-hololens-handtracking, soares2021accuracy-hololens-hand-tracking}), allowing participants to use their hand's gestures to move the box (Manipulate) and pop bubble (Select), as they would in the physical world. Pointing and clicking gestures are excluded from the \Direct conditions to maintain realism. Each gesture step flows into the next with tactile feedback (cursor or ray or color change) to the participants so that gestures are interpreted correctly by the system. Following HoloLens 2's near-interaction guidelines, primary objects (boxes and bubbles) are placed 35-50 cm from the participant's abdomen for ease of access and are ergonomically positioned to prevent muscle fatigue~\cite{gorilla-arm-effect-hansberger2017dispelling}. To reduce accidental selections, participants are taught specific hand gestures consistent with the standard HoloLens 2 hand gestures interaction paradigm~\cite{interactable-object}. Lastly, in Manipulate conditions, the box was placed on the table to provide haptic feedback.

\section{User Study}
\label{sec:study}
\begin{figure}[t]
    \centering
    \includegraphics[width=\columnwidth]{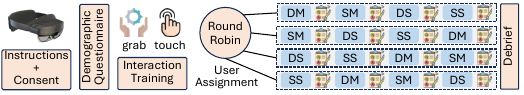}
    \vspace{-0.3cm}
    \caption{User study timeline consisting of pre-experiment procedures and users performing all four experimental conditions in one of the four orders devised using Latin square order. The four experimental blocks include \DM, \SM, \DS, \SSelect.}
    \vspace{-0.3cm}
    \label{fig:study-flow}
\end{figure}

\begin{figure}[t]
    \centering
    \includegraphics[width=.6\columnwidth]{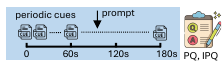}
    \vspace{-0.6cm}
    \caption{Individual experimental block.}
    \label{fig:block-flow}
\end{figure}

This section presents the details of our user study, experimental conditions, measures, and procedures.
\begin{table}[t]
\scriptsize
\caption{Demographic data and media usage (past 5 years) across all conditions and participants. The key for frequency: never/almost never; rarely ($<2$times); occasionally (a few times); frequently in the past;
frequently ($>2$times/month).}
\vspace{-0.15cm}
\label{tab:demographic}
\resizebox{\columnwidth}{!}{%
\begin{tabular}{l|l}
\hline
\textbf{demographics} & \textbf{\# participants} \\ \hline
\hline
gender                & 17 female; 33 male              \\ \hline
age                   & mean = 22.7 years (STD = 4.43)                             \\ \hline
\begin{tabular}[c]{@{}l@{}}frequency of \\ VR experience\end{tabular} &
 \begin{tabular}[c]{@{}l@{}}10 never used; 10 rarely; 20 occasionally;\\  5 frequently; 5 frequently in the past\end{tabular} \\ \hline
\begin{tabular}[c]{@{}l@{}}frequency of \\ AR/MR experience\end{tabular} &
 \begin{tabular}[c]{@{}l@{}}17 never used; 17 rarely; 12 occasionally;\\  4 frequently; 0 frequently in the past\end{tabular} \\ \hline
\begin{tabular}[c]{@{}l@{}}frequency of \\ Gaming\end{tabular} &
 \begin{tabular}[c]{@{}l@{}}5 never used; 3 rarely; 12 occasionally; \\ 25 frequently; 5 frequently in the past\end{tabular} \\ \hline
\end{tabular}%
}
\vspace{-0.6cm}
\end{table}

\vspace{-0.2cm}
\subsection{Participants}
\vspace{-0.1cm}
We conducted an a priori power analysis to determine the appropriate sample size for our within-subject four conditions (\DM, \SM, \DS, \SSelect) study for medium size effect ($0.25$)~\cite{cohen2013statistical-effect-size} with four repeated measures, an $\alpha = 0.05$, a power of $0.95$,  and nonsphericity correction ($\epsilon = 1$), the analysis recommended a minimum of 39 participants. 
We recruited 50 participants to account for attrition (33 male-identifying, 17 female-identifying) with a mean age of 21.86 years (ST = 3.39). All participants were recruited through the college campus. 
All participants provided written informed consent and received \$15 for their participation. All the participants had normal or corrected vision with contact lenses or glasses and the absence of motion sickness history. 
The demographic distribution and media usage of participants
can be seen in Table \ref{tab:demographic}. The study was approved by the UMASS Amherst IRB, Human Research Protection Office.

\vspace{-0.2cm}
\subsection{Material}
\vspace{-0.1cm}
We selected Hololens 2~\cite{hololens2} due to its direct view of the real world, a self-contained device with a holographic processing unit with a Qualcomm Snapdragon 850 CPU, featuring eye, spatial, and hand-tracking capabilities and dual displays with a $1440\times936$ pixels resolution with 110\textdegree FoV.
The VEs were developed using the Unity game engine and C\# programming language with the Unity XR Plug-ins, optimized for the Universal Windows Platform. Custom scripts were added to facilitate hand tracking and reaction time recording. 
\subsection{Experimental Task}
\vspace{-0.1cm}
In Section \S\ref{sec:exp-design}, we explain the experimental conditions and designs. Here, we will summarize the physical tasks (move, touch, grab, point), virtual object placements, and how they are ordered in the study session.
The study features four conditions to evaluate two interactions across two tasks (as shown in \figureautorefname~\ref{fig:teaser}). 
We designed a color-matching activity involving a bubble and four boxes for two Select task conditions. Participants stand as the scene is rendered before them, with bubbles positioned at their eye level and adjusted for height.
Every 3s (seconds), the bubble changes color, and every 9s, the box color changes.  
Participants must pop the bubble when its color matches the box's, complete 20 bubble-popping trials, and set the total phase duration to 3 minutes.
In the \Direct, participants pop bubbles by directly touching them, while in the \Symbolic, they use an air-tap gesture to interact with the boxes (\S\ref{sec:exp-design}). Here, the phase length is determined by time rather than the exact number of trials, ensuring a fixed session length across participants while maintaining consistency in the task structure.

For two Manipulate task conditions, we designed a box moving activity using a single virtual box. Participants are asked to sit in a chair positioned in front of a table and adjust the chair so they can comfortably reach the box. Every 5 seconds, the box turns green, signaling the participant to move the box to the side. Starting from a neutral position, they move the box to the right in one trial and then to the left in the following trial, continuing this alternating pattern. Participants perform 35 trials of moving the box within a phase that also lasts 3 minutes. The phase duration is again set by time, with the number of trials serving as a reference point for participants to maintain pace.
To prevent participants from moving the box too far, white tape marks were pasted on the table 8 inches apart to indicate the target areas. The distance of movement was empirically determined during the pilot study. In the \Direct interaction, participants use both hands to move the box with a real-world mimicking grab gesture directly. In contrast, in the \Symbolic interaction, they use a pinch gesture to manipulate hooks on the box (\S\ref{sec:exp-design}). The durations for both tasks were determined during pilot testing to balance participant engagement and avoid fatigue (\S\ref{sec:pilot}).

In all conditions, we measure the \emph{reaction time} as an interval between the time the color change occurred (cue) and the actual action taken by the participant (interaction). Reaction time was chosen as an objective performance metric based on prior findings linking faster responses to increased presence~\cite{chandio-tvcg-23, chandio-vr-24-human-factors}. This metric complements subjective presence scores, providing a real-time proxy for cognitive engagement. The feedback is explicit in all conditions: successful actions are marked by either the bubble popping or the box changing position, depending on the specific condition.
To build a comprehensive response time profile, we concurrently recorded timestamps of all events, including color changes, touch, movement, and transitions, in a separate thread. Additionally, we logged the distance each participant moved the box, identified which box was moved, and verified if it matched the highlighted box. This comparison determined the accuracy: a perfect match indicated the correct color pairing. While memory retention is not a focus of this study, we log this data to exclude response times for incorrect color matches, ensuring comparability across all participants and conditions.
\setlength{\tabcolsep}{3.5pt}
\begin{table*}[t]
   \centering
   \scriptsize
   \caption{Summary of descriptive results for presence score using PQ and IPQ questionnaires, prompt score,  and reaction time across experiments:  direct selection (\DS), symbolic selection (\SS), direct manipulation (\DM), and symbolic manipulation (\SM). \textbf{All entries are given as mean value (standard deviation)}. The subscales for PQ and IPQ questionnaires are Possibility to act (ACT), interface quality (IFQUAL), realism (REAL), possibility to examine (EXAM), self-evaluation of performance (EVAL), general presence (GP), spatial presence (SP), and involvement (INV).}
   \vspace{-0.1cm}
   \begin{tabular}{l|ccccc|cccc|c|c}
       \toprule
       \textbf{Exp.} & \multicolumn{5}{c}{\textbf{PQ} (1--7)} & \multicolumn{4}{c}{\textbf{IPQ} (1--7)} & \textbf{Prompt} & \textbf{Reaction Time}\\ \cline{2-11}
       
       \rule{0pt}{2.5ex}  & \textbf{ACT} & \textbf{IFQUAL} & \textbf{REAL} & \textbf{EXAM} & \textbf{EVAL} & \textbf{GP} & \textbf{SP} & \textbf{INV} & \textbf{REAL} & (1--7) & \textbf{(seconds)} \\
       \midrule
       \DS & 5.35 (1.07)     & 5.05 (1.19)   & 4.91 (1.34)   & 5.18 (1.34)   & 5.50 (1.42)   & 5.00 (1.30)   & 4.46 (0.79)   & 4.22 (0.96)   & 4.22 (0.97) & 5.05 (1.05) & 1.065 (0.43) \\
       
       \SS & 4.27 (1.10)     & 4.15 (1.22)   & 3.85 (1.01)   & 4.89 (1.24)   & 5.24 (0.96)   & 4.92 (0.91)   & 4.19 (0.62)   & 3.70 (0.69)   & 3.50 (1.19) & 4.55 (1.30) & 1.824 (0.21) \\
       
       \DM & 5.91 (0.74)     & 5.55 (0.99)   & 5.50 (0.83)   & 5.75 (0.94)   & 6.25 (0.94)   & 5.40 (1.36)   & 4.18 (0.45)   & 4.80 (0.68)   & 4.62 (0.84) & 5.23 (1.24) & 1.915 (0.54) \\
       
       \SM & 4.55 (0.96)     & 4.57 (0.91)   & 4.43 (0.80)   & 5.59 (1.04)   & 6.30 (0.75)   & 5.38 (1.39)   & 4.08 (0.52)   & 4.11 (0.62)   & 3.92 (1.18) & 5.08 (1.10) & 2.215 (0.71)\\
       \bottomrule 
   \end{tabular}
   \label{tab:descriptive-summary}
   \vspace{-0.2cm}
\end{table*}
\subsection{Measurements}
\vspace{-0.1cm}
We focus on presence and reaction time as the primary variables in this study to investigate how they relate to different interaction mechanisms and tasks, as outlined in our hypotheses (\S\ref{sec:hyp}).
We measured presence using two post-experience questionnaires: the Witmer and Singer Presence Questionnaire (PQ)\cite{ws} and the Igroup Presence Questionnaire (IPQ)\cite{ipq,ipq-2}. The PQ evaluates factors such as the possibility to act and examine, realism, self-evaluation, and interface quality. The IPQ measures three factors: the sense of physically being in the VE, realism, and involvement (crucial for evaluating interactions).
The presence scores are derived from 33 items (14 IPQ, 19 PQ) on a 7-point scale with no modifications. Additionally, we included a prompt question, "How connected do you feel to this virtual world?" on a 7-point scale, created by~\cite{chandio-vr-24-human-factors}, to measure presence within the scene between trials in all four conditions and to align with reaction time. Presence is the dependent variable in \textbf{H1, H1.1, H2, H2.1}, where interaction type (independent variable) and task type (moderating variable in \textbf{H2}) are expected to affect the participants' sense of presence. 

Our software recorded reaction time on the HoloLens2 in milliseconds. We collected 20 reaction time measurements for each participant for the Select conditions and 35 for the Manipulate conditions. Reaction time is the dependent variable in \textbf{H3}, where we examine its relationship with presence and how it varies based on interaction type (independent variable in \textbf{H3.1}) and the magnitude of this effect (examined in \textbf{H3.2}). As the task phases were time-based rather than trial-completion-based, the analysis accounted for any variability in the number of completed trials when calculating average reaction times.

\vspace{-0.2cm}
\subsection{Pilot Study}
\label{sec:pilot}
\vspace{-0.1cm}
Before the main study, we ran a qualitative pilot with five participants to fine-tune the task parameters to minimize biases and enhance study reliability. By employing a think-aloud protocol, we collected insights on participant comfort and observations without steering preferences toward specific parameters. This included testing different intervals for color changes (2, 3, 5, 10, and 15 seconds) and session duration (2 to 10 minutes). We also explored varying distances for box placement (5 to 20 inches) and assessed participant feedback on physical discomfort, color visibility, and object transparency. Feedback indicated a preference for a consistent size scale of boxes and bubbles; transparent bubbles were generally less favored. The pilot study's data helped refine the main study's design, ensuring tasks were ergonomically viable and technologically functional.
Using the desktop application to establish consistent baseline times across all conditions, we assessed reaction times and accuracy (color matching) in Select conditions.

\vspace{-0.2cm}
\subsection{Procedure}
\label{sec:procedure}
\vspace{-0.1cm}
The study procedure is shown in Figure ~\ref{fig:study-flow}. Participants began by reviewing and signing the consent form, then providing demographic information like gender, age, and familiarity with AR/VR/gaming. They were briefed on the study conditions (experiments and questionnaires), scene applications, headset features, interactions to be performed (instruction on how to perform), appropriate hand poses, prompt questions, visual stimuli, actions they need to perform, and feedback they will receive as a result of successful interaction. In the briefing, participants are instructed to prioritize task accuracy in each condition, emphasizing correct responses over speed.

In within-subjects designs, managing order effects is essential to measure the experimental conditions' impact accurately. To handle this, we used a balanced Latin square design~\cite{preece1991latin-square} to randomize the order of the four conditions across the participants. Each condition was placed in each position an equal number of times and directly followed others equally. The 4x4 Latin square pattern, repeatedly multiple times in a round-robin manner, enabled proper distribution:
\{\DM, \SM, \DS, \SSelect \}, 
\{ \SM, \DS, \SSelect, \DM\}, 
\{\DS, \SSelect, \DM, \SM\}, 
\{\SSelect, \DM, \SM, \DS\}
This approach made sure that each condition occupied every position once per set of four participants. 
In all conditions, participants wore the headset and performed two successful interactions (bubble popping and box moving) to familiarize themselves before each condition. These practice interactions were recorded but excluded from the analysis. 
As shown in \figureautorefname~\ref{fig:block-flow}, periodic cues are generated (3s in Select and 5s in Manipulate), and participants answered a prompt question halfway through each condition (1.5 minutes in). Participants removed the headset after three minutes, completing all trials (35 in Manipulate conditions and 20 in Select conditions) and answering the prompt question. After each condition, they filled out presence questionnaires and responded to an open-ended question (general comments used for qualitative analysis in \S\ref{sec:thematic-analayis}). After completing all four conditions, we conducted a small debriefing session where we answered their questions. The study session, including the initial briefing, four conditions, questionnaires, and debrief, took about 45 minutes.

\section{Results}
\label{sec:eval}
This section presents the results from our experiments using a within-subject study design when measuring presence using presence questionnaires (subjective) and reaction time (systematic). 
We conducted our analysis using the  \texttt{scipy} library in Python.

\subsection{Statistical Analysis}
\autoref{tab:descriptive-summary} summarizes the descriptive results from our statistical analysis, where we aggregated presence scores from questionnaires, prompt scores, and reaction time values. 
For the PQ and IPQ presence questionnaires, we reported values for all of the subscales: Possibility to act (ACT), interface quality (IFQUAL), realism (REAL), possibility to examine (EXAM), self-evaluation of performance (EVAL), general presence (GP), spatial presence (SP), and involvement (INV). 
We report results for all experimental scenarios separately.
For the ACT subscale, \DM ($5.91 \pm 0.74$) and \DS ($5.35 \pm 1.07$) show higher scores than \SM ($4.55 \pm 0.96$) and \SS ($4.27 \pm 1.10$), indicating stronger perceived control in \Direct. Similarly, IFQUAL scores are higher in \DM ($5.55 \pm 0.99$) and \DS ($5.05 \pm 1.19$) compared to Symbolic. For PQ-REAL, \DM ($5.50 \pm 0.83$) shows higher scores than \SM ($4.43 \pm 0.80$), and \DS ($4.91 \pm 1.34$) exceeds \SS ($3.85 \pm 1.01$).
For the EXAM, \DM ($5.75 \pm 0.94$) scores higher than \SM ($5.59 \pm 1.04$), and \DS ($5.18 \pm 1.34$) is higher than \SS ($4.89 \pm 1.24$), indicating greater capability to explore \Direct. In the EVAL, \SM ($6.30 \pm 0.75$) shows the highest score across all conditions, followed by \DM ($6.25 \pm 0.94$).
For the IPQ-REAL, \DM ($4.62 \pm 0.84$) shows higher scores than \SM ($3.92 \pm 1.18$), and \DS ($4.22 \pm 0.97$) exceeds \SS ($3.50 \pm 1.19$). GP is highest for \DM ($5.40 \pm 1.36$), and INV is higher for DM ($4.80 \pm 0.68$) compared to \SM ($4.11 \pm 0.62$).
Overall, \Direct led to higher scores in all subscales compared to \Symbolic across both Select and Manipulate.

The prompt score, a real-time measure of presence, shows small variations across conditions, with \DM showing the highest average score ($5.23 \pm 1.24$) and \SS\xspace the lowest ($4.55 \pm 1.30$).
The reaction time results indicate that participants generally responded faster in direct interaction tasks. In the \DS, the average reaction time was shorter ($1.065 \pm 0.43$ sec) compared to the \SS condition ($1.824 \pm 0.21$ sec). Similarly, \DM had a faster average reaction time ($1.915 \pm 0.54$ sec) compared to \SM ($2.215 \pm 0.71$ sec).

We used Levene's homogeneity test, which held true for all dependent variables. 
The Shapiro-Wilk test found that the aggregate questionnaire, individual questionnaires, prompt score, and reaction time were normally distributed. 
Among the different subscales, self-evaluation of performance (EVAL), possibility to examine (EXAM), and general presence (GP) failed the normality test, while other subscales were found to be normally distributed. We do not evaluate the impact of our dependent variables on the three aforementioned subscales, in addition to spatial presence (SP), as summary statistics show that interactions and tasks did not significantly impact these subscales.
\begin{table}[ht]
    \centering
    \scriptsize
    \caption{Summary of the ANOVAs' results using questionnaires and reaction time as a presence measure. [\texttt{\textbf{i}:interaction} and \texttt{\textbf{t}:task}] }
    \vspace{-0.1cm}    
    \begin{tabular}{llcccccc}
        \toprule
       \textbf{Source} & \textbf{Factor} & \textbf{SS} & \textbf{df} & \textbf{MS} &\textbf{F-stat} & \textbf{\emph{p}} & $\eta^2$  \\ 
        \midrule
        \textbf{Questionnaire} & & & \\\cline{1-1} 
        
        \rule{0pt}{2.5ex}Main effects & \texttt{i} &	22.87	&	1	&	22.87&  57.38 & $<$0.05 & 0.207 \\
        
        & \texttt{t} &	9.00	&	1	&	9.00& 22.57 & $<$0.05 & 0.082  \\ \cline{1-1}
        
        \rule{0pt}{2.5ex}Cross effect & \texttt{i:t} &	0.001	&	1	&	0.001& 0.002 & 0.967 & 0.002 \\ \cline{1-1}

        \rule{0pt}{2.5ex}\textbf{ACT} & & & \\\cline{1-1} 
        
        \rule{0pt}{2.5ex} Main effects & \texttt{i} &75.95&	1	&	75.95& 78.79 & $<$0.05 & 0.276 \\
        
        & \texttt{t} &	9.57	&	1	&	9.57 & 9.927 & $<$0.05 & 0.035  \\ \cline{1-1}

        \rule{0pt}{2.5ex}\textbf{IFQUAL} & & & \\\cline{1-1} 
        
        \rule{0pt}{2.5ex} Main effects & \texttt{i} &	44.18	&	1	&	44.18& 36.72 & $<$0.05 & 0.152 \\
        
        & \texttt{t} &	10.58	&	1	&	10.58& 8.794 & $<$0.05 & 0.036  \\ \cline{1-1}

        \rule{0pt}{2.5ex}\textbf{PQ-REAL} & & & \\\cline{1-1} 
        
        \rule{0pt}{2.5ex} Main effects & \texttt{i} &	56.94	&	1	&	56.94& 53.85 & $<$0.05 & 0.203 \\
        
        & \texttt{t}&	16.90	&	1	&	16.90 & 15.98 & $<$0.05 & 0.060  \\ \cline{1-1}

        \rule{0pt}{2.5ex}\textbf{INV} & & & \\\cline{1-1} 
        
        \rule{0pt}{2.5ex} Main effects & \texttt{i} &	18.30	&	1	&	18.30&  31.93 & $<$0.05 & 0.128 \\
        
        & \texttt{t} &	12.25	&	1	&	12.25& 21.37 & $<$0.05 & 0.085  \\ \cline{1-1}

        \rule{0pt}{2.5ex}\textbf{IPQ-REAL} & & & \\\cline{1-1} 
        
        \rule{0pt}{2.5ex} Main effects & \texttt{i} &	25.03	&	1	&	25.03& 22.00 & $<$0.05 & 0.098 \\
        
        & \texttt{t}&	8.30	&	1	&	8.30& 7.307 & $<$0.05 & 0.032  \\ \hline

        \rule{0pt}{2.5ex}\textbf{Prompt Score} & & & \\\cline{1-1} 
        
        \rule{0pt}{2.5ex} Main effects & \texttt{i} &	5.41	&	1	&	5.41& 3.899 & 0.0497 & 0.019 \\
        
        & \texttt{t} &	6.37	&	1	&	6.37& 4.591 & 0.0333 & 0.022  \\ \cline{1-1}
        
        \rule{0pt}{2.5ex} Cross effect & \texttt{i:t} &	1.53	&	1	&	1.53& 1.103 & 0.2948 & 0.005 \\ \hline
        
        \rule{0pt}{2.5ex}\textbf{Reaction Time} & & & \\\cline{1-1} 
        
        \rule{0pt}{2.5ex} Main effects & \texttt{i} &	14.03	&	1	&	14.03& 54.43 & $<$0.05 & 0.163 \\
        
        & \texttt{t} &	19.24	&	1	&	19.24& 74.62 & $<$0.05 & 0.223  \\ \cline{1-1}
        
        \rule{0pt}{2.5ex} Cross effect & \texttt{i:t} &	2.62	&	1	&	2.62& 1.028 & 0.016 & 0.003 \\ 
        
        \bottomrule 
    \end{tabular}
    \label{tab:anova-results}
    \vspace{-0.6cm}
\end{table}

\autoref{tab:anova-results}  presents the results of the two-way repeated measures ANOVA, using combined questionnaire scores, individual subscores, prompt scores, and reaction time as the dependent variables. The analysis reveals that the main effect of interaction type is statistically significant across several subscales and dependent variables.
The interaction type shows a significant effect within the study's constraints ($F = 78.79, p < 0.05, \eta^2 = 0.276$), indicating that interaction type explains 27.6\% variance in ACT  subscale for the tested conditions. This finding highlights that interaction (\Direct vs. \Symbolic) substantially influences participants' perception of their ability to act within the environment.
The IFQUAL subscale shows a significant main effect of interaction type ($F = 36.72, p < 0.05, \eta^2 = 0.152$), indicating 15.2\% variance. This suggests that the quality of the interaction interface is influenced by whether the interaction is \Direct and \Symbolic.
For PQ-REAL, the interaction type effect is significant ($F = 53.85, p < 0.05, \eta^2 = 0.203$), indicating that interaction type accounts for 20.3\% of the variance in the perceived realism of the VE. In INV, the interaction type effect is significant, with $F = 31.93, p < 0.05, \eta^2 = 0.128$, showing that interaction type explains 12.8\% of the variance in involvement scores.
The IPQ-REAL subscale also exhibits a significant main effect of interaction type ($F = 22.00, p < 0.05, \eta^2 = 0.098$), meaning that interaction type indicates 9.8\% of the variance in this measure of realism.
For the prompt score, which captures participants' real-time sense of presence, the effect of interaction type is marginally significant, with $F = 3.899, p = 0.0497, \eta^2 = 0.019$, explaining 1.9\% of the variance.
The interaction type significantly affects reaction time ($F = 54.43, p < 0.05, \eta^2 = 0.163$), influencing participant reaction time in tested conditions.

\subsubsection{Task Type Effects}
The main effect of task type is statistically significant for several subscales. For the Possibility to Act (ACT) subscale, task type shows an effect with $F = 9.927, p < 0.05, \eta^2 = 0.035$. This indicates that the type of task (Select vs. Manipulate) influences participants' perception of their ability to act in the virtual environment.
For IFQUAL, task type shows a significant effect, with $F = 8.794, p < 0.05, \eta^2 = 0.036$. This suggests that task type impacts how participants evaluate the quality of the interface.
For PQ-REAL, task type shows a significant effect with $F = 15.98, p < 0.05, \eta^2 = 0.060$, 6\% of the realism scores variance.
The INV exhibits a significant task type effect, with $F = 21.37, p < 0.05, \eta^2 = 0.085$, indicating that task type influences participants' level of involvement in the environment.
For IPQ-REAL, task type shows a significant effect, with $F = 7.307, p < 0.05, \eta^2 = 0.032$. Regarding the prompt score, task type exerts a small but statistically significant effect, with $F = 4.591, p < 0.05, \eta^2 = 0.022$, indicating 2.2\% variance.
For reaction time, task type has the most substantial effect, with $F = 74.62, p < 0.05, \eta^2 = 0.223$, showing that the type of task significantly influences how quickly participants respond.

\subsubsection{Cross Effects}
The interaction effect between interaction type and task type is generally not significant for most dependent variables. However, there is a small but statistically significant interaction effect for reaction time, with $F = 1.028, p = 0.016, \eta^2 = 0.003$. This suggests that the combination of interaction type and task type has a minimal, yet notable, influence on reaction time.

\subsubsection{Correlation}
In~\autoref{fig:reaction_time_presence}, the relationship between presence scores and reaction time is visualized, with each plot representing one of the four experimental conditions. Pearson’s correlation coefficients across tested conditions showed a modest negative correlation between presence and reaction time, supporting the relationship between interaction type, task type, and user performance in the context of \Manipulate and \Select tasks, with values ranging from -0.42 in \DM to -0.63 in \SM with linear regression models fit, illustrating the inverse relationship between presence and reaction time across conditions.

 \begin{figure}[t]
     \centering
     \includegraphics[width=.8\columnwidth]{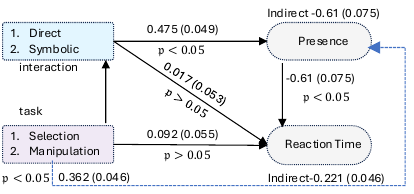}
     \vspace{-0.2cm}
     \caption{Mediation analysis showing the direct and indirect effects of interaction and task type on reaction time, with presence as a mediator. Arrows indicate relationships with coefficients, standard errors, and p-values. Significant indirect effects are shown for interaction and task types $ (p< 0.05)$.
}
     \label{fig:mediation}
 \end{figure}
 
\subsection{Mediation Analysis}
\begin{figure*}[t]
    \centering
    \begin{tabular}{cccc}
    \includegraphics[width=0.23\textwidth]{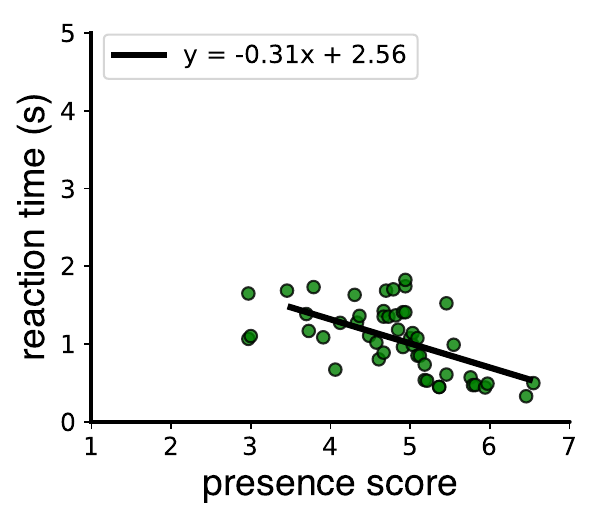} &
    \includegraphics[width=0.23\textwidth]{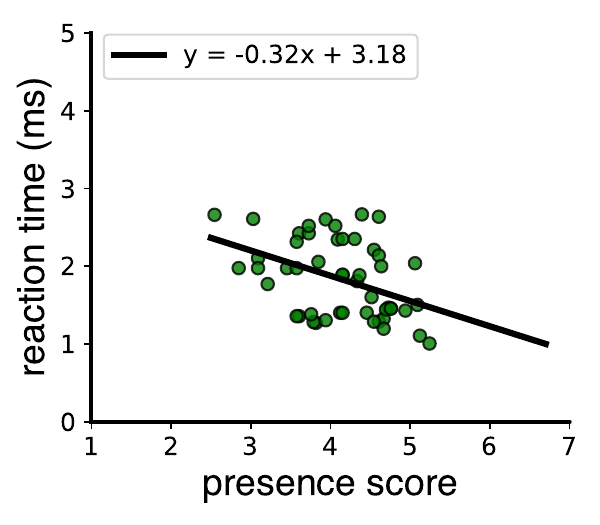} &
    \includegraphics[width=0.23\textwidth]{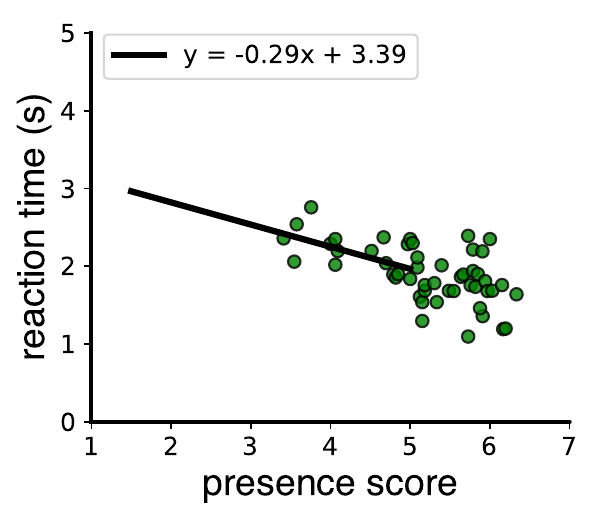} &
    \includegraphics[width=0.23\textwidth]{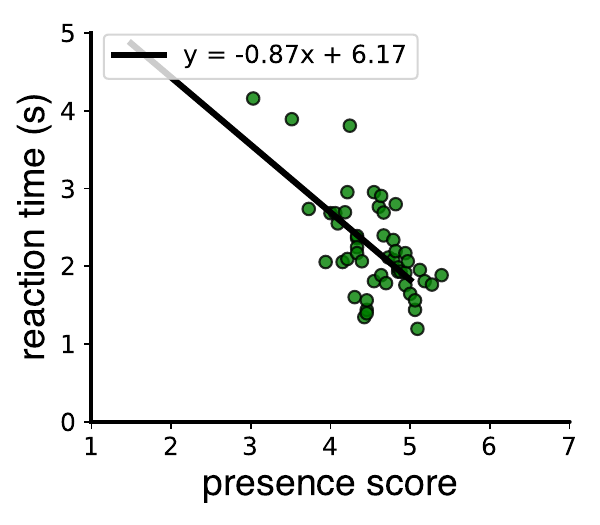} \vspace{-0.05cm} \\
    (a) Direct Selection (\DS) & (b) Symbolic Selection (\SS) & (c) Direct Manipulation (\DM) & (d) Symbolic Manipulation (\SM) \\
    \end{tabular}
    \vspace{-0.3cm}
    \caption{\emph{Presence vs. Reaction Time: Presence decreases as reaction time increases. Reaction time and presence also show a modest correlation with Pearson’s correlation coefficients: overall (-0.54), \DS (-0.59), \SS (-0.42), \DM (-0.59), and \SM (-0.63). Each green circle represents a study participant. The black line is the linear regression fit for the data.}}
    \vspace{-0.5cm}
    \label{fig:reaction_time_presence}
\end{figure*}

We conducted the mediation analysis (~\autoref{fig:mediation}) to understand the relationship between interaction, task, and reaction time, with presence as a mediator, by using a bootstrapping approach with 1000 resamples with 95\% confidence intervals (CI).
The analysis revealed a positive effect of interaction type on presence, with a coefficient of $0.475$ ($p < 0.05$, $CI [0.380, 0.573]$). This suggests that \Direct resulted in higher levels of presence than \Symbolic interactions, indicating a strong relationship between interaction type and presence. For the task type, we also examined its effect on presence. The results showed that Select generally resulted in higher levels of presence than Manipulate tasks. 
For Task Type, the effect on presence was slightly smaller, with a coefficient of $0.362$ ($p < 0.05$, $CI [0.245, 0.485]$), suggesting that Select tasks led to higher levels of presence than Manipulate tasks. The next phase of our analysis focused on the relationship between presence and reaction time. We found that Presence significantly negatively affected reaction time in the studied conditions, with a coefficient of $-0.610 (p < 0.05$,  $CI [0.245, 0.485]$), suggesting that as presence increased, reaction time became faster (also in ~\autoref{fig:reaction_time_presence}). However, this effect may be influenced by user familiarity and task type~\cite{freeman2000focus}, though disentangling these factors is beyond the scope of our study.

The indirect effect of interaction type on reaction time, mediated by Presence, was calculated to be $-0.290 (p < 0.05$, $CI [-0.398, -0.190]$). This suggests that the improvement in reaction time associated with interaction type was largely mediated by the sense of presence.
We also calculated the indirect effect of task type on reaction time, again mediated by Presence. This indirect effect was $-0.221 (p < 0.05$, $CI [-0.310, -0.130]$), showing that Select improved presence, leading to faster reaction times. 
The direct effect of interaction type on reaction time, controlling for presence, is found to be $0.017 (p > 0.05$, $CI [-0.087, 0.122]$) 0.017, which is small and not statistically significant. This indicates that once presence is accounted for, interaction type had little direct influence on reaction time. This suggests that the effect of interaction type on reaction time is primarily mediated by presence.
The direct effect of task type on reaction time, controlling for presence, is $0.092 (p > 0.05$, $CI [-0.015, 0.199]$). This indicates a small positive effect of Select on reaction time, meaning that this direct effect is not statistically significant. Thus, the mediator explains the relationship between tasks and reaction time.

\subsection{Control Measures}
The participants' familiarity with AR/MR, VR, and video games was not evenly distributed, as shown in~\autoref{tab:demographic}. 
Due to this uneven distribution, an extensive statistical analysis on familiarity was not possible, and these control measures were not considered further. 
As discussed in the study procedure (\S\ref{sec:procedure}), the participants were assigned one of the four experiment orders based on the Latin square pattern to manage the order effect. 
To confirm that the experiment order did not impact our outcomes, we conducted ANOVA of presence score, prompt score, and reaction time against the experiment order. 
To do that, an aggregate value using the sum across all experiments was computed for each user. 
The four groups of users were compared using experiment order as the independent variable. 
Our results suggested that experiment order did not significantly impact the presence scores ($\text{F}$=0.79, $\text{p}$=0.54), prompt score ($\text{F}$=1.34, $\text{p}$=0.13), and reaction time ($\text{F}$=2.513, $\text{p}$=0.09).

\subsection{Interpretation of Results}
In this section, we interpret the effects identified in our statistical analysis and explore the reasons for their manifestation. 
Our study used the overall change in presence as the interaction mechanism changed as the first check; the relationship between presence and reaction time would not be observed if the presence did not change. 
We observed a significant difference in overall presence, which is found to be statistically significant (\autoref{tab:anova-results}). 
Therefore, we accept \textbf{H1}.
We also found that the change in the interaction mechanism changed the overall presence, and the individual subscales showed consistent statistical significance.
The interaction mechanisms differed in enabling users to act, the interface quality, how realistic they felt, and the level of involvement they needed to work. 
As a result, the interaction type is the dominant independent factor that significantly impacted the presence score across subscales. 
Furthermore, as the direct interactions were more involved than the \Symbolic, the presence score for \DS and \DM is higher (\autoref{tab:descriptive-summary}) and statistically significant, and mediation analysis also supports \textbf{H1.1} as the indirect effect of interaction type on reaction time, mediated by presence, shows that \Direct result in higher presence. Thus, \textbf{H1.1} can be accepted. 
The different tasks differed in the level of involvement they expected from the participant; it is expected that changing a task would impact presence. However, we also expected that the task's nature might impact the interaction type's effect on presence; for example, the participant may find a given interaction mechanism better suited to one task than the other. 
However, our results suggested that while task impacts presence, the effect of the two interaction types on presence does not change. We did not observe an interaction effect at the overall questionnaire and subscales levels (omitted for brevity). 
Therefore, \textbf{H2} is rejected.
Our next hypothesis took a more specific approach to the cross-over between task and interaction, stating that there will be a change in presence if the interaction changes. This is observed, and \textbf{H2.1} can be accepted.

The first four hypotheses focused on the relationship between presence score, task type, and interaction type. 
We next examined the relationship between these factors and reaction time.
Understandably, reaction time is naturally impacted by the type of interaction. 
For example, a simple touch interaction would take much less time than a manipulation interaction that requires a more careful analysis of the virtual and physical worlds. 
However, the reaction time is not just the function of the interaction type, but also the task type. 
Therefore, it is important to analyze that the effect of interaction type is consistent irrespective of the task type. 
In our study, we examined the correlation between presence and reaction time across different experimental conditions. As shown in~\autoref{fig:reaction_time_presence}, a moderate to high negative correlation is observed between presence scores and reaction time, with Pearson’s correlation coefficients: overall (-0.54), \DS (-0.59), \SS (-0.42), \DM (-0.59), and \SM (-0.63). These results confirm the negative relationship between presence and reaction time, supporting \textbf{H3}.
Additionally, as shown in~\autoref{tab:anova-results}, the impact of interaction type on the reaction time is statistically significant, and \textbf{H3.1} is accepted. Additionally, similar to its effect on presence, the task type also impacted reaction time. 
However, reaction time is the only measure and the dependent variable on which we observed the cross-effect between interaction type and task type. 
The score in presence questionnaires is the same, and the interaction effect explains most of the variance in the observed score. 
However, in reaction time, both tasks take different amounts of time; therefore, the task type and interaction type explain a significant variance. 
In our statistical analysis, the effect of task type also shows that these systemic measures, such as reaction time, depend on the task at hand in addition to the user's level of presence.  In mediation analysis \textbf{H3, H3.1} is supported that changes in interaction type influence reaction time via presence. Additionally, \textbf{H3.2} is not fully validated, showing that task type impacts reaction time, primarily through its effect on presence. The direct effect of interaction type on reaction time, however, is not significant.
As a result, it demonstrates a cross-effect in addition to the main effects. 
Therefore, we partially accept \textbf{H3.2}.

\subsection{Thematic Analysis}
\label{sec:thematic-analayis}
While our quantitative results demonstrate that different conditions affected presence scores and reaction times, we wanted to investigate the factors contributing to these changes. To do this, we disaggregated the answers to the open-ended question and subscales: 
We do not present any results for general and spatial presence-related comments as they are highly correlated with overall scores.

\noindent \textbf{Involvement:}
As both quantitative and qualitative results suggested, transitioning between different experimental conditions often caused participants to feel a heightened sense of involvement. Comments such as ``Being able to control the VE directly was engrossing" and ``I felt really part of the virtual world when the interactions were responsive" were common in \DM and a few in \DS.

\noindent \textbf{Realism:}
Similarly, the shift to conditions \DM and \DS was noted by the participants who commented, ``..objects looked almost tangible," and "The realistic response of the environment to my actions added to the fun". This suggests that enhancements in visual and interaction realism contribute to the sense of presence.

\noindent \textbf{Interface Quality:}
Feedback on interface quality often highlighted the limitations of the field of view and control responsiveness. For example, one participant noted, ``..small field of view limited my interaction," while another said, ``Sometimes the gestures did not register, which pulled me out of the experience" (reported in Select). These comments align with observations where interface limitations negatively impact user experience.

\noindent \textbf{Comfort:}
The physical comfort of using the headset also played a crucial role in the presence experience. Participants frequently mentioned comfort issues, stating, ``The headset felt heavy after a while," or "I experienced some eye strain during the last few trials." 

The thematic analysis revealed consistent themes across conditions, especially regarding the impact of technical features on the presence. Enhancements in realism and interaction quality were positively received, while limitations in interactions and physical discomfort were areas of concern. These insights from thematic analysis are supported by the quantitative results from the subscales, providing a comprehensive understanding of the factors influencing presence in VE.

\section{Discussion}
\label{sec:disc}

Our findings revealed significant interaction and task type effects on presence scores, as indicated by our two-way repeated measures ANOVA. The interaction type showed a higher influence on the presence scores across all subscales, while task type had a more pronounced effect on reaction time. Notably, the interaction between task and interaction type had a negligible impact, suggesting the influence of interaction type on presence and reaction time across different tasks.

\vspace{0.1cm}
\noindent \textbf{Correlation between Presence and Reaction Time.}
The correlation between presence and reaction time was modestly negative, ranging from -0.42 to -0.63 across different scenarios. Previous studies in immersive VR environments have also observed similar negative correlations between presence and reaction time, suggesting that enhanced presence reduces reaction time. This aligns with our findings, reinforcing the relationship across both \Direct and \Symbolic interactions within our study. This correlation is visualized in \autoref{fig:reaction_time_presence}, with linear regression lines illustrating the relationship between these measures. The trend indicates that as presence increases, reaction time decreases, suggesting that higher levels of presence in MR can lead to faster responses.

\vspace{0.1cm}
\noindent \textbf{Implication of Task Design.}
Participants reported higher presence during Select tasks than Manipulate tasks, likely due to the brief, frequent interactions reinforcing a sense of agency and success. Selection tasks provide frequent feedback that aligns with user expectations in MR, enhancing presence. In contrast, Manipulate tasks require sustained motor coordination, potentially shifting focus to task execution. Prior research~\cite{response-activation-2019} suggests continuous motor tasks can increase cognitive load and reduce presence. Future work could explore how interaction frequency impacts presence across varied tasks.

 \vspace{0.1cm}
\noindent \textbf{Interpretation of Hypotheses.}

Our results indicate that participants experienced heightened presence with direct interactions compared to symbolic interactions in the studied tasks. At the same time, this supports H1 and H2; we acknowledge that the outcomes are constrained by the experimental scope and task design.
However, this observation also reflects the intuitive nature of direct interactions that closely mimic real-world actions, facilitating a stronger engagement with the VE.
H3 was not supported, highlighting that direct interaction's superior efficacy in enhancing presence is not contingent on the task type. 
The findings suggest a preference for direct interactions over symbolic ones in MR; however, this preference may vary with different task complexities and interaction goals.
H2.1 was confirmed, underscoring that interaction type changes are sufficient to significantly alter the sense of presence. This finding is crucial for designers of MR interfaces, as it emphasizes the importance of choosing appropriate interactions to enhance the overall experience.
H3, H3.1, and H3.2 highlight the relationships between interaction types, reaction times, and presence, confirming that interaction types affect the presence and reaction times.

\noindent\textbf{Theoretical and Practical Implications.}
These results have multifaceted implications. Theoretically, they contribute to understanding how different interactions influence user experience in MR, offering insights into the psychological and physiological aspects of presence. Practically, these findings can guide the development of more effective MR applications, where the choice of interaction type can be tailored to improve both the sense of presence and performance efficiency.

\section{Limitations and Future Work}
\label{sec:future}

This section briefly discusses the limitations of 
our study. We also outline potential future directions for this work.

One challenge we faced was using presence questionnaires originally designed for VR environments in an MR context. Due to the lack of validated MR-specific presence measures, we used VR questionnaires, which offer the most established and reliable way to assess presence. Additionally, the questionnaire score can indicate the level of presence a user is experiencing. However, since reaction time also depends on the task, just looking at it does not provide enough information about the level of presence. At present, reaction time is best suited for measuring the change in presence rather than the absolute level of presence. 
Further studies should explore whether the effect of task type on reaction time can be modeled, as this may enable the use of reaction time as a standalone absolute measure of presence. 
In our study, we focused on two interaction types (\Direct and \Symbolic) and two tasks (Select and Manipulate) for clarity and control, allowing for a focused examination of foundational MR interactions before exploring more complex variations in future research.
Our study focused on reaction time as a performance measure due to its real-time reflection of cognitive and motor effort during MR interactions. However, future work could incorporate complementary metrics, such as accuracy, task completion time, and subjective workload, to capture a more comprehensive view of performance, particularly for more complex tasks.
Lastly, further research could explore the long-term effects of these interactions on user adaptation and learning.
Analyzing prolonged exposure to varied interactions will reveal how users adapt over time, potentially developing new strategies or behaviors. This research could uncover patterns in the correlation between sustained exposure and presence scores and how extended exposure influences reaction times. Exploring multisensory feedback integration can significantly enhance immersion and presence. Researchers could examine how these additional sensory channels influence the relationship between presence and reaction time by incorporating auditory, haptic, and olfactory cues. VE that allow multi-user interactions provide a unique opportunity to study social presence. Researching how users perceive others in shared virtual spaces and how collaboration impacts interactions, overall presence, and reaction time could be an exciting avenue. 

\section{Conclusion}
\label{sec:conclusion}
In conclusion, this study provides a comprehensive analysis of how interaction and task types influence presence and reaction time, offering valuable insights for enhancing user experience in MR environments. The confirmation of H1 and H2 highlights the immersive potential of direct interactions for tasks requiring high immersion compared to symbolic interactions. Future work should examine additional symbolic interactions to provide broader generalizability. 
The study confirms that changes in interaction type alone can meaningfully alter presence and reaction time. This finding has practical implications for MR developers, highlighting the importance of careful interaction designs to enhance user experience. 
In summary, these findings expand our understanding of MR interactions and point to reaction time as a promising metric for assessing presence in future MR applications, warranting further exploration.

\acknowledgments{%
We thank the anonymous IEEE VR 2025 reviewers for their insightful comments and feedback. The research reported in this paper was sponsored by the National Science Foundation under award 2237485.
}

\bibliographystyle{abbrv-doi-hyperref}
\bibliography{paper}

\end{document}